\documentclass[reprint,superscriptaddress,amsmath,amssymb,aps,twocolumn]{revtex4-1}
\usepackage{color}
\usepackage{bm,bbm}
\usepackage{epsfig}
\usepackage{amsmath}
\usepackage{amsfonts}
\usepackage{amssymb}
\usepackage{multirow}
\usepackage{amsbsy}
\usepackage{array}
\usepackage{diagbox}
\usepackage{tikz}
\usepackage{extarrows}
\usepackage{tabularx}
\usepackage{appendix}
\usepackage{txfonts}
\usepackage[normalem]{ulem}
\usepackage{graphicx}
\usepackage{url}
\usepackage[colorlinks=true, urlcolor=blue, linkcolor=blue, citecolor=blue, pdftex]{hyperref}
\DeclareUnicodeCharacter{2212}{-}
\begin{document}

\newcommand{\fig}[2]{\includegraphics[width=#1]{#2}}
\newcommand{\la}{{\langle}}
\newcommand{\ra}{{\rangle}}
\newcommand{\dg}{{\dagger}}
\newcommand{\upa}{{\uparrow}}
\newcommand{\dna}{{\downarrow}}
\newcommand{\ab}{{\alpha\beta}}
\newcommand{\ias}{{i\alpha\sigma}}
\newcommand{\ibs}{{i\beta\sigma}}
\newcommand{\hH}{\hat{H}}
\newcommand{\hn}{\hat{n}}
\newcommand{\hc}{{\hat{\chi}}}
\newcommand{\hU}{{\hat{U}}}
\newcommand{\hV}{{\hat{V}}}
\newcommand{\br}{{\bf r}}
\newcommand{\bk}{{{\bf k}}}
\newcommand{\bq}{{{\bf q}}}
\def\gsim{~\rlap{$>$}{\lower 1.0ex\hbox{$\sim$}}}
\setlength{\unitlength}{1mm}
\newcommand{{\vhf}}{$\chi^\text{v}_f$}
\newcommand{{\vhd}}{$\chi^\text{v}_d$}
\newcommand{{\vpd}}{$\eta^\text{v}_d$}
\newcommand{{\ved}}{$\epsilon^\text{v}_d$}
\newcommand{{\vved}}{$\varepsilon^\text{v}_d$}

\title{Orbital-Selective $d$-wave Superconductivity in the Two-Band $t$-$J$ Model: \\Possible Applications to La$_3$Ni$_2$O$_7$}

\author{Zhan Wang}
\affiliation{Beijing National Laboratory for Condensed Matter Physics and Institute of Physics, \\Chinese Academy of Sciences, Beijing 100190, China}

\author{Kun Jiang}
\email{jiangkun@iphy.ac.cn}
\affiliation{Beijing National Laboratory for Condensed Matter Physics and Institute of Physics, \\Chinese Academy of Sciences, Beijing 100190, China}
\affiliation{School of Physical Sciences, University of Chinese Academy of Sciences, Beijing 100190, China}

\author{Fu-Chun Zhang}
\email{fuchun@ucas.ac.cn}
\affiliation{School of Physical Science and Technology, ShanghaiTech University, Shanghai 201210, China}
\affiliation{Kavli Institute for Theoretical Sciences and School of Quantum, University of Chinese Academy of Sciences,	Beijing, 100190, China}
    
\author{Hui-Ke Jin}
\email{jinhk@shanghaitech.edu.cn}
\affiliation{School of Physical Science and Technology, ShanghaiTech University, Shanghai 201210, China}
\date{\today}

\begin{abstract}
{
We investigate superconductivity in a two-band $t$-$J$ model consisting of an itinerant orbital (orbital-0) and a quasi-localized orbital (orbital-1) using variational Monte Carlo. A robust orbital-selective $d$-wave superconducting state is found to emerge exclusively from the itinerant orbital. An analysis of the superexchange energy hierarchy shows that the quasi-localized orbital-1 competes with superconductivity by favoring local inter-orbital bound states, which act as energy defects and disrupt phase coherence. Consistently, the superconducting order parameter is monotonically suppressed as the occupancy of orbital-1 increases. Motivated by superconductivity in nickelate La$_3$Ni$_2$O$_7$, these results highlight the essential role of multi-orbital physics beyond the single-band $t$-$J$ framework and point to a concrete route to enhance $T_c$: suppressing the involvement of localized $d_{z^2}$-derived orbitals.
}
\end{abstract}

\maketitle

{\color{blue} \em Introduction}---
High-$T_c$ superconductivity (SC), spurred by the discovery of cuprates, remains a central challenge in condensed matter physics, driving decades of research into strongly correlated electron systems~\cite{2004Zhang, 2006Wen, 2015Zaanen}. The single-band $t$-$J$ model has become a canonical framework for addressing this problem~\cite{1988Rice,2006Wen, 2008Fukuyama}. It provides an effective low-energy description of the cuprates' intrinsic three-band structure, a simplification justified by the formation of the robust Zhang-Rice singlet~\cite{1988Rice}. Crucially, extensive numerical studies, particularly Variational Monte Carlo (VMC) simulations~\cite{1988Shibaa, 1988Shibab, 1989Gros, 1999Ogata, 1991Lhuillier, 1996Ogata, 2001Trivedi, 2003Himeda, 2004Trivedi, 2004Lee}, have consistently demonstrated that the SC ground state hosts a robust $d$-wave pairing \cite{Tsuei_RevModPhys.72.969}, establishing it as a cornerstone for the theory of cuprates. Nevertheless, results from more sophisticated numerical calculations indicate that the stability of long-range SC order in this model remains controversial~\cite{White1998,Corboz2014,Zheng2017,Qin2020}.

In contrast, the two-band $t$-$J$ model---the most natural extension beyond the single-band paradigm---remains far less explored. This implies a historical lack of material platforms requiring such a description. 
Usually, orbital effects are suppressed by the large level splitting. For example, in cuprates, the large Jahn-Teller distortion lifts the $e_g$ orbital degeneracy, resulting in a local electronic configuration with predominant $3d_{x^2-y^2}$ character. Exceptions arise only in specific cases, such as LaNiO$_3$/LaMO$_3$ superlattices ~\cite{2008Khaliullin}, highly overdoped cuprates~\cite{2016Xue, 2018Jiang} or Ba$_2$CuO$_{3+\delta}$ under high pressure~\cite{2019Jin}, where the two $e_g$ orbitals become nearly degenerate. This leaves a fundamental question unresolved: how does the inclusion of the second active orbital impact the well-established $d$-wave SC state?

This situation has been changed by the recently discovered high-$T_c$ SC in the nickelate family~\cite{2019Hwang, 2023Wangc, 2025Hwangc, chengjg_crystal, 2025Chenb, 2024Zhaoa, 2025Qi, 2024Xiangb, 2025Chene}, particularly in the bilayer ($n=2$) Ruddlesden-Popper (RP) compounds exemplified by La$_3$Ni$_2$O$_7$~\cite{2023Wangc, 2025Hwangc, chengjg_crystal, 2025Chenb, chengjg_poly, 2024Yuan,  2024Feng, 2024Chenc, 2024Mathon,  2025Wangi, 2025Wangg, 2025Wang, 2025Maoa, 2025Hwang, 2025Hwanga, 2025Hwangb, 2025Shen, 2025Xuea, 2025Zhanga, 2025Chenc, 2025He, 2025Wangd,  2025Nie, 2025Niea, 2025Tsukazaki}. The RP structure consists of corner-sharing NiO$_6$ octahedra, producing a local crystal field similar to that of the cuprates. A pivotal distinction, however, arises from the nickel valence: in the RP nickelates, it is governed by the layer number~$n$, evolving from Ni$^{2+}$ ($3d^8$) to Ni$^{3+}$ ($3d^7$)~\cite{2023Wangc,2025Hwangc,2025Chene}. This results in a tunable multi-orbital electronic environment, a feature that fundamentally differentiates them from the single-orbital cuprates. Consequently, the discovery of high-$T_c$ SC in RP nickelates transforms the two-band $t$-$J$ model from a theoretical curiosity into a framework of immediate relevance. The intricate interplay among orbital degrees of freedom, interlayer coupling, and strong correlations in La$_3$Ni$_2$O$_7$ is the key to addressing the nature of SC, stimulating great theoretical interest~\cite{2023Yao,
2023Wangd, 2023Zhango, 2023Yangb, 2023Eremin, 2023Si, 2024Zhangf, 2025Xianga, 2024Wua, 2024Kurokia, GangSu_PhysRevLett.132.036502, 2024Wangi, 2024Wehling, 2024Hub-dft, 2024Hirschfelda, 2025Jiang-dft, 2025Hua-dft, 2025Wangl, 2025Hua, 2025Hug, 2025Zhange, 2025Zhangh, 2025Li, 2025Chen, 2025Kuroki, 2025Chaloupka, 2026Kontani, 2026Raghu, 2026Jiang}

In this letter, we analyze the impact of the second orbital on SC within a generic two-band $t$-$J$ model. The model is defined by two orbitals with distinct mobilities: an itinerant orbital (orbital-0) and a quasi-localized orbital (orbital-1). Our analysis of the model's energy hierarchy reveals a fundamental principle, namely, the second orbital is universally detrimental to SC. Our central result, supported by VMC calculations, is that the system develops robust {\em orbital-selective $d$-wave pairing}, which arises exclusively from the itinerant orbital-0. In contrast, the localized orbital-1 exhibits a vanishing pairing amplitude and acts effectively as a source of ``energy defects'' that disrupt the coherent condensate. This general mechanism provides a framework for understanding multi-orbital materials and yields a prediction for systems like the high-$T_c$ nickelates (e.g., La$_3$Ni$_2$O$_7$): any tuning parameter that suppresses the participation of the second orbital will enhance $T_c$.

\begin{figure*}
    \centering
    \includegraphics[width=0.95\linewidth]{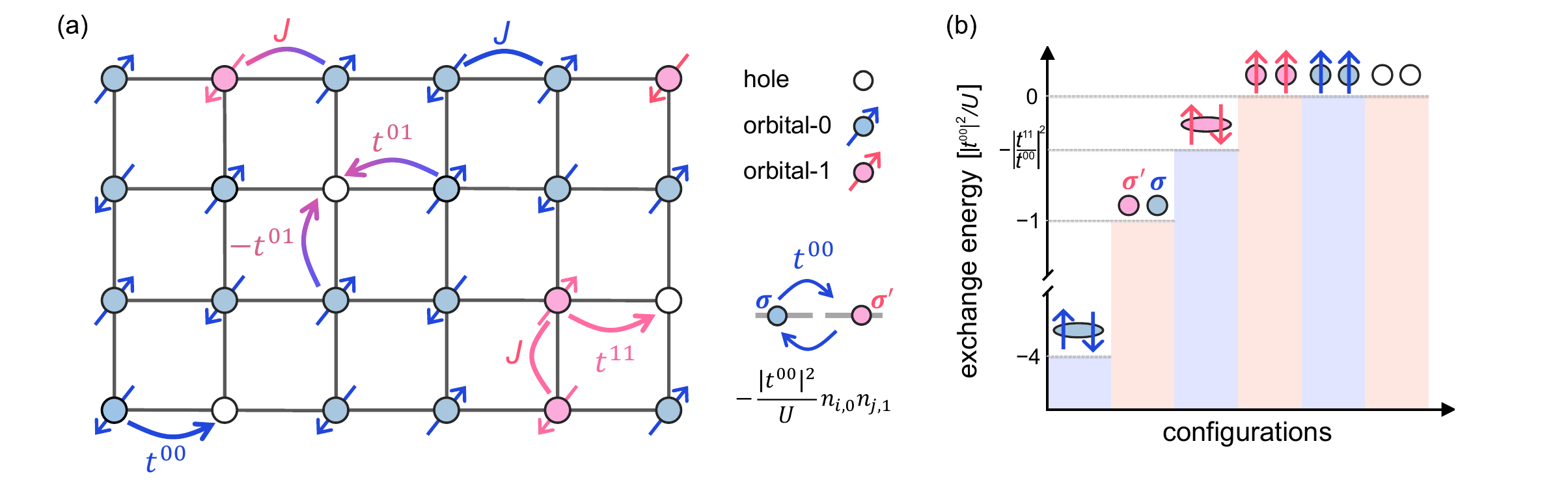}
    \caption{\textbf{Schematic of the 2-band $t$-$J$ model defined in Eq.~\eqref{eq:H_total} and the local superexchange energy hierarchy.} \textbf{(a)} Left panel: Schematic of the two-band $t$–$J$ model. Electrons in orbital-0 (orbital-1) are shown in blue (pink), and holes are indicated by empty circles. The hopping processes for $t^{00}, t^{11}$, and $t^{01}$ are indicated by arrows between nearest-neighboring sites. Note that the inter-orbital hopping $t^{01}$ takes opposite signs along the $x$- and $y$-directions. The superexchange $J\sim t^{\alpha\beta}t^{\beta^\prime\alpha'}/U$ is a rank-4 tensor  depending on the orbital indices. Right panel: The intra-orbital-0 superexchange process yields the inter-orbital density-density term in Eq.~\eqref{eq:Hex0}. This interaction is mediated solely by $t^{00}$ and is spin-independent, as indicated by the general spin indices $\sigma$ and $\sigma^\prime$. \textbf{(b)} Schematic energy hierarchy of superexchange interactions between two neighboring sites in the regime $t^{00}\gg t^{01},t^{11}$. Ellipse with spin up and down denotes a spin singlet. The energy level of $E=-1$ corresponds to the configuration in which different orbitals are occupied at neighboring sites, irrespective of their spins.}
    \label{fig:schematic}
\end{figure*}

{\color{blue} \em  Model Hamiltonian}---As the starting point, we employ a two-band $t$-$J$ model on a square lattice, as illustrated schematically in Fig.~\ref{fig:schematic}(a). This model is derived from a Kugel-Khomskii model at quarter-filling with a large on-site Hubbard $U$~\cite{1973KK, 2025Zhange}. The model Hamiltonian is
\begin{equation}
\mathcal{H} = P_G\mathcal{H}_{\rm kin}P_G + \mathcal{H}_{\rm ex},    \label{eq:H_total}
\end{equation} 
where $P_G$ is the Gutzwiller projector forbidding double (and also higher) occupancy. The kinetic term, defined on the nearest-neighbor bonds $\langle ij \rangle$, reads 
\begin{equation}
\mathcal{H}_{\rm kin} = - \sum_{\langle ij\rangle,s} \sum_{\alpha,\alpha'} t_{ij}^{\alpha\alpha'} \left(c^\dagger_{i\alpha s} c_{j\alpha's} + \text{h.c.}\right) - \mu \sum_{i,\alpha,s} n_{i\alpha s}.
\label{eq:Hkn}
\end{equation}
Here, $c^\dagger_{j\alpha s}$ creates an electron with spin $s$ at site $j$ in orbital $\alpha$, and $n_{j\alpha s}=c^\dagger_{j\alpha s}c_{j\alpha s}$. For practical reasons, we consider the two orbitals as the $e_g$ orbitals on the $3d$ shell. Specifically, we attribute the orbital index $\alpha=0$ to the itinerant $d_{x^2-y^2}$-like band with larger isotropic hopping $t^{00}$, and $\alpha=1$ the quasi-localized $d_{z^2}$-like band with a much smaller isotropic hopping $t^{11}$. Note that the wavefunction of $d_{x^2-y^2}$ has opposite signs along the $x$- and $y$-axes while $d_{z^2}$-like orbital-1 is symmetric in the plane. This results in a sign change for inter-orbital hopping: $t_{i,i+\hat{\mathbf{x}}}^{01} =-t_{i,i+\hat{\mathbf{y}}}^{01}= t^{01}$. The exchange term $\mathcal{H}_{\rm ex}$, derived from the large-$U$ limit (neglecting Hund's coupling)~\cite{2025Zhange}, governs the spin and orbital fluctuations:
\begin{equation}
\begin{split}
\mathcal{H}_{\rm ex}=&\sum_{\langle ij\rangle}\sum_{\alpha\alpha^\prime\beta\beta^\prime} J^{\alpha\beta\beta'\alpha'}_{ij}\left\{
{\bf S}_{i,\alpha\alpha^\prime} \cdot {\bf S}_{j,\beta^\prime\beta} - \frac{1}{4}n_{i,\alpha\alpha^\prime} n_{j, \beta^\prime\beta}\right. \\ 
&\qquad\left. +\frac{\left(-1\right)^{\delta_{\beta\beta'}}}{4}  
n_{i,\alpha\alpha'}n_{j,\bar{\beta}\bar{\beta}'}+\frac{\left(-1\right)^{\delta_{\alpha\alpha'}}}{4}n_{i,\bar{\alpha}'\bar{\alpha}}n_{j,\beta'\beta}\right\},
\end{split}
\label{eq:Hex}
\end{equation}
with coupling $J^{\alpha\beta\beta'\alpha'}_{ij} = 4t_{ij}^{\alpha\beta}(t_{ji}^{\beta^\prime\alpha'})^*/U$. We use generalized spin $\mathbf{S}_{i,\alpha\beta} = \frac{1}{2} c_{i\alpha}^\dagger \bm{\sigma} c_{i\beta}$ and density $n_{i,\alpha\beta} = c_{i\alpha}^\dagger c_{i\beta}$ operators, where $c^\dagger_{i\alpha}=(c^\dagger_{i\alpha\uparrow}, c^\dagger_{i\alpha\downarrow})$. Note that the notation of generalized operators $\mathbf{S}_{i,\alpha\beta}$ and $n_{i,\alpha\beta}$ differs from the widely used orbital pseudospin notations, see Appendix~\ref{app:spin_notation}.

The full Hamiltonian $\mathcal{H}$ hosts a rich phenomenology, highlighted by an SU(4) symmetry at the point $t^{01}=0$ and $t^{11}=t^{00}$. While a comprehensive exploration of the phase diagram is certainly desirable, here we focus on quarter-filling ($n=1$) to provide crucial insights directly relevant to SC in La$_3$Ni$_2$O$_7$. Note that at quarter-filling and with orbital-1 inactive, this system reduces to an effective half-filling single-band $t$-$J$ model for orbital-0 only, which is known for hosting $d$-wave SC upon proper doping. However, realistic nickelate materials are inherently a true two-orbital system with  $t^{11} \approx 0.2 t^{00}$ and $t^{01} \approx 0.5 t^{00}$~\cite{2025Jiang}. The primary goal of this paper is to investigate how the presence of orbital-1 competes with and modifies the $d$-wave pairing tendencies of the dominant orbital-0.

{\color{blue} \em  SC suppressed by energy defects}---With one electron per site, charge fluctuations are frozen in $\mathcal{H}$, and the low-energy physics is governed {\em solely} by $\mathcal{H}_{\rm ex}$. We consider the limit where only orbital-$0$ is itinerant ($t^{01}, t^{11} \rightarrow 0$). 
In this limit, all effective interactions arise from virtual hopping processes involving orbital-0. The resulting effective Hamiltonian $\mathcal{H}_{\rm ex}'$ is:
\begin{equation}
\mathcal{H}_{\rm ex}' = \frac{|t^{00}|^2}{U} \sum_{\langle ij \rangle} (4\mathbf{S}_{i,0} \cdot \mathbf{S}_{j,0} -n_{i,0} n_{j,0} - n_{i,0} n_{j,1} - n_{i,1}n_{j,0}),
\label{eq:Hex0}
\end{equation}
where we denote $\mathbf{S}_{i,\alpha} \equiv \mathbf{S}_{i,\alpha\alpha}$ and $n_{i,\alpha} \equiv n_{i,\alpha\alpha}$. This Hamiltonian reveals two distinct types of interactions originating from the same superexchange mechanism. The first two terms constitute the usual single-band Heisenberg antiferromagnetic (AFM) superexchange for orbital-0. Crucially, as depicted in Fig.~\ref{fig:schematic}(a), the presence of orbital-1 also enables {\em spin-independent} superexchange processes involving orbital-0, leading to an inter-orbital density-density interaction. 

The presence of the quasi-localized orbital-1 establishes a distinct energy hierarchy for local electronic configurations within  $\mathcal{H}_{\text{ex}}$, see Fig.~\ref{fig:schematic}(b). The lowest energy scale is set by the more itinerant orbital-0 electrons, forming singlets to gain an energy $-4(t^{00})^2/U$ for a two-site system, and $-2.34(t^{00})^2/U$ per bond in the 2D limit~\cite{Sandvik1997}, with the energy from $n_{i,0}n_{j,0}$ accounted for. The key departure from the single-band picture emerges at the next energy level: a strong inter-orbital density-density attraction $-n_{i,0} n_{j,1}$ of order $\sim(t^{00})^2/U$ binds an orbital-1 electron to an orbital-0 electron irrespective of the spin orientations. This binding energetically dominates over the spin-dependent superexchange associated with the orbital-1, such as terms $\sim (t^{11})^2/U$ and $\sim (t^{01})^2/U$. 

Upon appropriate hole-doping, it is well-established that the Heisenberg AFM spin correlations in orbital-0 typically yield $d$-wave SC. This picture is directly disrupted by a finite $t^{01}$. As shown in Fig.~\ref{fig:schematic}(a), a small $t^{01}$ induces a finite occupancy $\langle n_1 \rangle \propto{}|t^{01}/t^{00}|^2$ in orbital-1. Electrons in orbital-1, via the aforementioned inter-orbital attraction, tend to form ``bound states'' with their orbital-0 partners, effectively acting as a spin-inert ``defect''. This process effectively sequesters the participating orbital-0 electrons, and meanwhile, prevents orbital-1 from developing its own coherent AFM spin correlations. Moreover, the low occupancy density in orbital-1 inhibits a coherent condensate. Consequently, the pairing correlations indicate an orbital-selective nature driven primarily by intra-orbital-0 pairings. Rather than fostering an additional pairing channel, the localized orbital-1 serves primarily to suppress the $d$-wave superconductivity within orbital-0.

\begin{figure}
    \centering
    \includegraphics[width=\linewidth]{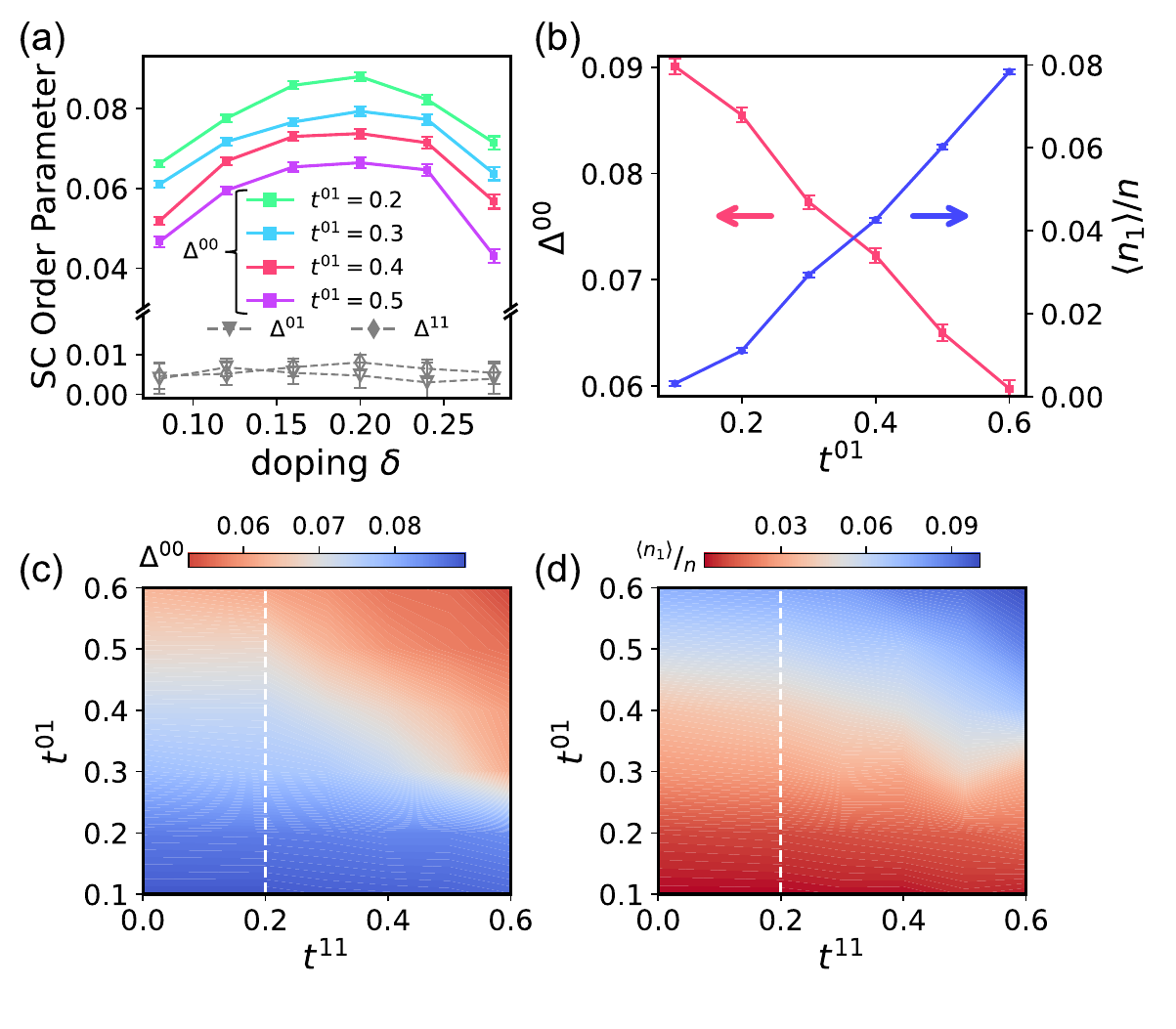}
    \caption{\textbf{VMC results on SC order parameters and orbital occupation number.} (a) SC order parameter as a function of doping $\delta$, obtained under various $t^{01}$ with $t^{11}=0.2$ fixed. For $\Delta^{00}$, the error bar is comparable to the marker size. The pairing order parameter on other channels is about $\Delta^{11} \sim \Delta^{01} \approx 0.005$ for $t^{01}=0.5$, which is very close to their errorbars. (b) Pairing order parameter $\Delta^{00}$ and the occupation ratio $\langle n_1\rangle/n$, obtained as a function of $t^{01}$, with $t^{11}=0.2$ fixed. (c) and (d), density plot for $\Delta^{00}$ and $\langle n_1\rangle/n$ as functions of $t^{01}$ and $t^{11}$. The white dashed line corresponds to the parameter used in (b). Other parameters: $t^{00}=1$, $U=8$ and lattice size of $L\times L$ with $L=20$. Doping $\delta=0.16$ $(n=0.84)$ in (b-d).}
    \label{fig:vmc}
\end{figure}

{\color{blue} \em Variational Monte Carlo results}---To quantitatively investigate the influence of orbital-1 on the SC order, we numerically simulate the ground state of the two-band $t$-$J$ model $\mathcal{H}$. Our approach utilizes a Gutzwiller projected wavefunction~\cite{1989Gros}, $|\psi\rangle=P_G|\psi_{\rm MF}\rangle$, which serves as the trial wavefunction for $\mathcal{H}$. Here, $|\psi_{\rm MF}\rangle$ is derived from a mean-field Bogoliubov-de Gennes (BdG) Hamiltonian. The variational parameters of this BdG Hamiltonian, including nearest-neighbor hopping amplitudes and $d$-wave pairing parameters, are optimized by minimizing the variational energy $E=\langle \psi|\mathcal{H}|\psi\rangle/\langle\psi|\psi\rangle$ using the VMC method~\cite{vmc_sorella_1, vmc_sorella_2}. While the VMC approach allows for the exploration of various trial wavefunction ans\"atze to identify the true ground state, our primary interest in this work lies in quantitatively investigating how the celebrated $d$-wave SC evolves with the additional orbital in the $t$-$J$ model. Consequently, this study predominantly presents results focusing on the $d$-wave SC. It is important to note that we also explored other potential pairing symmetries, such as $s$-wave and $s+id$-wave. However, these alternative states consistently yielded significantly higher variational energies, confirming $d$-wave symmetry as the energetically dominant pairing channel under the current model parameters. 

The simulations are performed on an $L\times L$ lattice with periodic boundary conditions for various doping ratios $\delta$, corresponding to a total of $\delta{}L^2$ doped holes. We set the energy scale with $t^{00}=1$ and the Hubbard interaction to $U=8$. The inter-orbital hopping parameters, $t^{01}$ and $t^{11}$, are varied in the range of $[0.1, 0.6]$. For a detailed description of the wavefunction construction and simulation methodology, please refer to Appendix~\ref{methods}.

To investigate the evolution of the $d$-wave SC state with varying doping levels and model parameters, we calculate the equal-time pair-pair correlation function $\Phi^{\alpha\beta}_{\mathbf{x}}(r)$ along the $\hat{\mathbf{x}}$-direction:
\begin{equation}
    \Phi_{\mathbf{x}}^{\alpha\beta}(r)=\left\langle \left(D_{i,j}^{\alpha\beta}\right)^\dagger D_{i+r\hat{\mathbf{x}},j+r\hat{\mathbf{x}}}^{\alpha\beta} \right\rangle.
\end{equation}
Here $D_{i,j}^{\alpha\beta}=c_{i\alpha\uparrow}c_{j\beta\downarrow}-c_{i\alpha\downarrow}c_{j\beta\uparrow}$ denotes the singlet pairing operator on the nearest neighbor bond $\langle ij\rangle$. The correlation $\Phi^{\alpha\beta}_{\mathbf{y}}(r)$ along the $\hat{\mathbf{y}}$-direction is computed analogously and exhibits similar behavior. We observe that the correlation $\Phi_{\mathbf{x}}^{\alpha\beta}$ quickly saturates to a plateau value for distance $r>3$. Therefore, the long-range SC order parameter $\Delta^{\alpha\beta}$ is extracted from this saturation value as $\Delta^{\alpha\beta}= (\bar{\Phi}^{\alpha\beta})^{1/2}$, where $\bar{\Phi}^{\alpha\beta}$ is the average of $\Phi^{\alpha\beta}_{\mathbf{x}}(r)$ over the plateau region $3<r<L/2$. 

The first key result is that the intra-orbital correlation $\Phi_{\mathbf{x}}^{00}$ is the sole dominant component, while all other pairing correlations are negligible; see Fig.~\ref{fig:vmc}(a). 
It demonstrates that the superconductivity is orbital-selective, where coherent SC pairing predominantly occurs within orbital-0.
Consistent with our previous arguments, the non-zero variational pairing parameters in the other channels merely manifest as a pseudogap, which fails to establish long-range pairing correlations.

Consistent with the results for single-band $t$-$J$ model~\cite{1988Shibab}, $\Delta^{00}$ also manifests a dome-like dependence on $\delta$. As shown in Fig.~\ref{fig:vmc}(a), this dome is centered at an optimal doping of approximately $\delta \approx 0.16 \text{--} 0.2$, with the precise location dependent on the specific model parameters.

A second central result is that the SC order parameter $\Delta^{00}$ is suppressed by the inter-orbital hopping $t^{01}$ across the entire doping range, see Fig.~\ref{fig:vmc}(a). This stems from enhanced band hybridization between the two orbitals, which in turn drives the transfer of electrons from orbital-0 to orbital-1. This charge redistribution is detrimental to $d$-wave pairing; as we have argued, the presence of occupancy in orbital-1 hinders the formation of coherent $d$-wave Cooper pairs within the orbital-0 background. To provide quantitative evidence, we compute the orbital occupancy $\langle n_\alpha\rangle = L^{-2} \sum_{i}\langle n_{i,\alpha}\rangle$ for various $t^{01}$, with the total density fixed at $n=\langle n_0\rangle + \langle n_1\rangle = 1-\delta$. Our numerical results directly validate this picture. As illustrated in Fig.~\ref{fig:vmc}(b), increasing $t^{01}$ leads to a suppression of $\Delta^{00}$ accompanied by a steady increase in the occupation ratio $\langle n_1\rangle/n$ of orbital-1. This strong correlation provides compelling evidence for our theory. 

The intra-orbital hopping $t^{11}$ provides an alternative pathway for suppressing the superconducting order parameter. By enlarging the bandwidth of orbital-1, increasing $t^{11}$ enhances its occupancy and consequently depletes the electron population of orbital-0. This effect is clearly demonstrated in Figs.~\ref{fig:vmc} (c) and (d) for a doping of $\delta=0.16$: both the SC order parameter and the orbital-0 occupancy are suppressed as either $t^{01}$ or $t^{11}$ is increased. This confirms the general principle that any parameter promoting charge transfer away from orbital-0 is detrimental to the stability of the $d$-wave SC state.

One might expect that the increase in $t^{01}$ or $t^{11}$ could foster a competing inter-orbital SC state. However, our analysis suggests this is not the case. We find that even for significant hopping values, such as $t^{01}, t^{11} \sim 0.5t^{00}$, the occupancy of orbital-1 remains relatively small, e.g., $n_1/n \lesssim 10\%$. 
Such a low occupancy density is insufficient to establish a phase-coherent condensate of Cooper pairs. Therefore, even if the subleading interactions are attractive and, in principle, allow for inter-orbital pairing, the resulting SC order remains orbital-selective, as robust phase coherence develops only within the intra-orbital channel $\Delta^{00}$. 
Any other potential pairing channels remain confined to a pseudogap-like regime.

Instead of promoting a new condensate, we argue that the enhanced coupling to orbital-1 tends to form local inter-orbital bound states; see Fig.~\ref{fig:schematic}(b). This finding points to a fundamentally orbital-selective 
nature for the SC, despite the two-band framework of the $t$-$J$ model. Our conclusion is also supported by the symmetric-limit calculation ($t^{11} = t^{00}$, $t^{01}>0$), where we find the optimized pairing variational parameter vanishes entirely. Thus, the second orbital acts not as a partner for superconductivity, but as a detrimental competing channel that suppresses the condensate anchored in orbital-0.

\begin{figure}
    \centering
    \includegraphics[width=\linewidth]{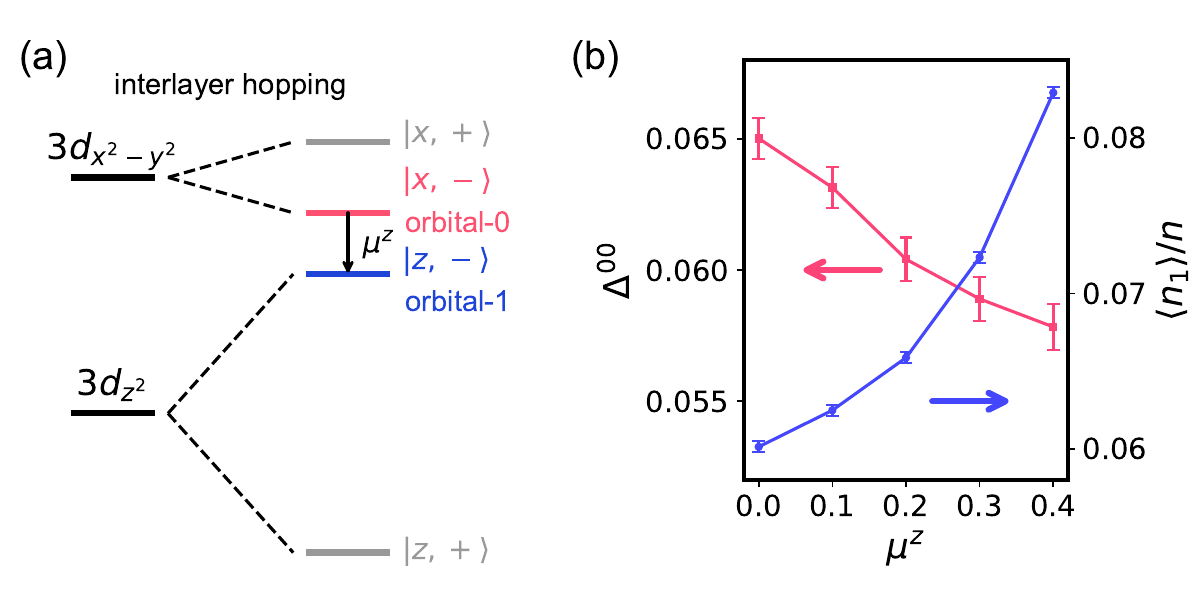}
    \caption{\textbf{Local electronic structure and $\mu_z$ effects on superconductivity.} (a) The local electronic structure of La$_3$Ni$_2$O$_7$. The two sets of $e_g$ orbitals split into molecular orbitals due to interlayer hoppings. The sign $\pm$ dictates the interlayer symmetry of the molecular orbitals. The antibonding orbital $|z,-\rangle$ and the bonding orbital $|x,-\rangle$ are nearly degenerate with a small energy separation denoted by $\mu^z$. (b) SC pairing $\Delta^{00}$ and $\langle n_1\rangle/n$ as functions of $\mu^z$. The energy unit is $t^{00}$. The parameters are $t^{00}=1$, $t^{01}=0.5$, $t^{11}=0.2$, $U=8$, doping $\delta=0.16$ and lattice size $L\times L$ with $L=20$.}
    \label{fig:muz}
\end{figure}

{\color{blue}\em Application to La$_3$Ni$_2$O$_7$.}---Our results offer key insights into the pairing mechanism of La$_3$Ni$_2$O$_7$ and suggest a possible route to higher $T_c$. To establish the model's relevance to La$_3$Ni$_2$O$_7$, we begin with its low-energy electronic structure, which is dominated by the Ni-$e_g$ orbitals ($d_{x^2-y^2}$ and $d_{z^2}$). Recent calculations~\cite{2024Zhangf,2025Jiang} indicate that strong interlayer hopping via $d_{z^2}$ orbitals hybridizes these atomic states into interlayer molecular orbitals of symmetric and antisymmetric characters. A low-energy effective theory can be constructed based on the local orbital configuration shown in Fig.~\ref{fig:muz}(a). The low-lying symmetric orbital, $|z,+\rangle$, is fully occupied and inactive. The two active bands near the Fermi level mainly consist of antisymmetric orbitals, i.e., $|x,-\rangle$ and $|z,-\rangle$. Moreover, a small electronic population in the higher-energy symmetric $|x,+\rangle$ orbital introduces a self-doping effect~\cite{2025Zhange}. 

In the strong coupling limit, La$_3$Ni$_2$O$_7$ is described as a self-doped molecular Mott insulator near quarter-filling~\cite{2025Zhange}, whose low-energy effective Hamiltonian naturally reduces to Eq.~\eqref{eq:H_total}. In this model, the itinerant (orbital-0) and localized (orbital-1) states correspond to the two interlayer antisymmetric molecular orbitals, $|x,-\rangle$ and $|z,-\rangle$, respectively. Numerical calculations~\cite{2025Jiang} suggest hopping parameters of $t^{01} \approx 0.5t^{00}$ and $t^{11} \approx 0.2t^{00}$, which situates the system within the parameter regime relevant to our study. However, these two orbitals are not perfectly degenerate in realistic systems. This quasi-degeneracy is particularly fragile and sensitive to perturbations such as crystal distortions~\cite{2025Hwangc, 2025Hwangb}, chemical substitution of rare-earth ions~\cite{2025Nie, 2025Niea, 2025Zhanga}, and variations in oxygen content~\cite{2025Wangd}. Although a concrete relation between the lattice/chemical perturbations and the orbital splitting remains to be established, we phenomenologically mimic these effects by incorporating  
the onsite term $-\mu^z\sum_{i,s}n_{i,1,s}$ into the Hamiltonian in Eq.~\eqref{eq:Hex}. This term lowers the onsite energy of orbital-1 ($|z,-\rangle$), as shown in Fig.~\ref{fig:muz}(a). 

Using VMC, we calculate the SC pairing order parameter and corresponding orbital occupancy as a function of the energy splitting $\mu^z$. As shown in Fig.~\ref{fig:muz}(b), increasing $\mu^z$ lowers the energy of the less-itinerant $|z,-\rangle$ orbital, leading to a higher $n_1$, the occupation of the $|z,-\rangle$ orbital. Meanwhile, the SC pairing order parameter $\Delta^{00}$ is monotonically suppressed with increasing $\mu^z$, indicating that a larger energy splitting between the orbitals is detrimental to SC. Note that the pairing is orbital-selective, and is dominated by the channel within the $|x,-\rangle$ orbital, with all other pairing channels being orders of magnitude smaller. 

This result further corroborates our previous analysis. We have now shown that two distinct parameters, $\mu^z$ and $t^{01}$, both control the occupation of the less-itinerant orbital ($n_1$), and consistently demonstrate that populating this orbital suppresses the SC order. This provides a clear strategy for enhancing $T_c$ in La$_3$Ni$_2$O$_7$-like systems. Experimental efforts, such as applying specific structural perturbations, should aim to reduce the orbital energy splitting depicted in Fig.~\ref{fig:muz}(a). Specifically, this means raising the on-site energy of the $|z,-\rangle$ orbital to bring it closer to degeneracy with the more itinerant $|x,-\rangle$ orbital.

{\color{blue}\em Discussion}---In summary, we study the ground state of a two-band $t$-$J$ model with coexisting itinerant and quasi-localized orbitals. VMC reveals a robust {\em orbital-selective} $d$-wave superconducting state arising solely from the itinerant orbital. Analysis of the superexchange energy hierarchy demonstrates a fundamental competition between the usual intra-orbital AFM correlations and unusual inter-orbital density interactions. Crucially, the localized orbital-1 acts as a competitor: its occupancy promotes local inter-orbital binding states which act as defects, thereby disrupting the phase coherence and suppressing the SC order parameter $\Delta^{00}$.

These results provide a microscopic framework for superconductivity in nickelates such as La$_3$Ni$_2$O$_7$, highlighting the dual role of multi-orbital physics. While the localized $|z,-\rangle$ orbital is essential for realizing the strong-coupling electronic structure, it simultaneously weakens the pairing order parameter. Our work suggests that enhancing $T_c$ requires suppressing the participation of this orbital, which may be achieved via strain, chemical substitution, or interface engineering. More broadly, the two-band $t$-$J$ model offers a platform to explore competing magnetic and orbital orders in multi-orbital correlated systems.

{\em Acknowledgments}---We acknowledge the support by the National Natural Science Foundation of China (Grant NSFC-12494594, NSFC-12574150, NSFC-12174428, NSFC-12504180), the Ministry of Science and Technology (Grant No. 2022YFA1403900), the Chinese Academy of Sciences Project for Young Scientists in Basic Research (2022YSBR-048), the Innovation program for Quantum Science and Technology (Grant No. 2021ZD0302500), Chinese Academy of Sciences under contract No. JZHKYPT-2021-08, and the start-up funding from ShanghaiTech University.

\bibliographystyle{apsrev4-2}
\bibliography{KKtJ}

@article{Tsuei_RevModPhys.72.969,
  title = {Pairing symmetry in cuprate superconductors},
  author = {Tsuei, C. C. and Kirtley, J. R.},
  journal = {Review of Modern Physics},
  volume = {72},
  issue = {4},
  pages = {969--1016},
  numpages = {0},
  year = {2000},
  month = {Oct},
  publisher = {American Physical Society},
  doi = {10.1103/RevModPhys.72.969},
  url = {https://link.aps.org/doi/10.1103/RevModPhys.72.969}
}

@article{2015Zaanen,
    author = {Keimer, B. and Kivelson, S. A. and Norman, M. R. and Uchida, S. and Zaanen, J.},
    da = {2015/02/01},
    doi = {10.1038/nature14165},
    id = {Keimer2015},
    isbn = {1476-4687},
    journal = {Nature},
    number = {7538},
    pages = {179--186},
    title = {From quantum matter to high-temperature superconductivity in copper oxides},
    ty = {JOUR},
    volume = {518},
    year = {2015}
}

@article{2006Wen,
  title = {Doping a Mott insulator: Physics of high-temperature superconductivity},
  author = {Lee, Patrick A. and Nagaosa, Naoto and Wen, Xiao-Gang},
  journal = {Review of Modern Physics},
  volume = {78},
  issue = {1},
  pages = {17--85},
  numpages = {0},
  year = {2006},
  month = {Jan},
  publisher = {American Physical Society},
  doi = {10.1103/RevModPhys.78.17}
}

@article{2004Zhang,
doi = {10.1088/0953-8984/16/24/R02},
year = {2004},
month = {jun},
publisher = {},
volume = {16},
number = {24},
pages = {R755},
author = {P W Anderson and  P A Lee and  M Randeria and  T M Rice and  N Trivedi and  F C Zhang},
title = "{The physics behind high-temperature superconducting cuprates: the ‘plain vanilla’ version
of RVB}",
journal = {Journal of Physics: Condensed Matter}
}

@article{2008Fukuyama,
  title = {The t--{{J}} Model for the Oxide High-{{Tc}} Superconductors},
  author = {Ogata, Masao and Fukuyama, Hidetoshi},
  year = 2008,
  month = feb,
  journal = {Reports on Progress in Physics},
  volume = {71},
  number = {3},
  pages = {036501},
  issn = {0034-4885},
  doi = {10.1088/0034-4885/71/3/036501},
  urldate = {2025-11-07},
  langid = {english},
}

@article{1988Rice,
    title = "{Effective Hamiltonian for the superconducting Cu oxides}",
    author = {Zhang, F. C. and Rice, T. M.},
    journal = {Physical Review B},
    volume = {37},
    issue = {7},
    pages = {3759--3761},
    numpages = {0},
    year = {1988},
    month = {Mar},
    publisher = {American Physical Society},
    doi = {10.1103/PhysRevB.37.3759}
}

@article{1988Shibaa,
  title = {A Renormalised {{Hamiltonian}} Approach to a Resonant Valence Bond Wavefunction},
  author = {Zhang, F. C. and Gros, C. and Rice, T. M. and Shiba, H.},
  year = {1988},
  month = jun,
  journal = {Superconductor Science and Technology},
  volume = {1},
  number = {1},
  pages = {36},
  issn = {0953-2048},
  doi = {10.1088/0953-2048/1/1/009},
  urldate = {2025-09-09},
  langid = {english},
}

@article{1989Gros,
  title = {Physics of Projected Wavefunctions},
  author = {Gros, Claudius},
  year = {1989},
  month = jan,
  journal = {Annals of Physics},
  volume = {189},
  number = {1},
  pages = {53--88},
  issn = {0003-4916},
  doi = {10.1016/0003-4916(89)90077-8},
  urldate = {2024-11-05},
}

@article{2001Trivedi,
  title = {Projected {{Wave Functions}} and {{High Temperature Superconductivity}}},
  author = {Paramekanti, Arun and Randeria, Mohit and Trivedi, Nandini},
  year = {2001},
  month = nov,
  journal = {Physical Review Letters},
  volume = {87},
  number = {21},
  pages = {217002},
  issn = {0031-9007, 1079-7114},
  doi = {10.1103/PhysRevLett.87.217002},
  urldate = {2025-09-09},
  copyright = {http://link.aps.org/licenses/aps-default-license},
  langid = {english},
}

@article{2004Trivedi,
  title = {High- {{T}} c Superconductors: {{A}} Variational Theory of the Superconducting State},
  shorttitle = {High- {{T}} c Superconductors},
  author = {Paramekanti, Arun and Randeria, Mohit and Trivedi, Nandini},
  year = {2004},
  month = aug,
  journal = {Physical Review B},
  volume = {70},
  number = {5},
  pages = {054504},
  issn = {1098-0121, 1550-235X},
  doi = {10.1103/PhysRevB.70.054504},
  urldate = {2025-09-09},
  copyright = {http://link.aps.org/licenses/aps-default-license},
  langid = {english},
}

@article{1996Ogata,
  title = {Phase {{Diagram}} and {{Pairing Symmetry}} of the {{Two-Dimensional}} t- {{J Model}} by a {{Variation Theory}}},
  author = {Yokoyama, Hisatoshi and Ogata, Masao},
  year = {1996},
  month = nov,
  journal = {Journal of the Physical Society of Japan},
  volume = {65},
  number = {11},
  pages = {3615--3629},
  publisher = {The Physical Society of Japan},
  issn = {0031-9015},
  doi = {10.1143/JPSJ.65.3615},
  urldate = {2025-09-09},
}

@article{1991Lhuillier,
  title = {Phase Diagrams of the Two-Dimensional {{Hubbard}} and {\emph{t}} - {{{\emph{J}}}} Models by a Variational {{Monte Carlo}} Method},
  author = {Giamarchi, T. and Lhuillier, C.},
  year = {1991},
  month = jun,
  journal = {Physical Review B},
  volume = {43},
  number = {16},
  pages = {12943--12951},
  issn = {0163-1829, 1095-3795},
  doi = {10.1103/PhysRevB.43.12943},
  urldate = {2024-12-30},
  copyright = {http://link.aps.org/licenses/aps-default-license},
  langid = {english},
}

@article{1999Ogata,
  title = {Coexistence of d x 2 - y 2 Superconductivity and Antiferromagnetism in the Two-Dimensional t - {{J}} Model and Numerical Estimation of {{Gutzwiller}} Factors},
  author = {Himeda, A. and Ogata, M.},
  year = {1999},
  month = oct,
  journal = {Physical Review B},
  volume = {60},
  number = {14},
  pages = {R9935-R9938},
  issn = {0163-1829, 1095-3795},
  doi = {10.1103/PhysRevB.60.R9935},
  urldate = {2025-09-09},
  copyright = {http://link.aps.org/licenses/aps-default-license},
  langid = {english},
}

@article{2003Himeda,
  title = {Superconductivity and {{Antiferromagnetism}} in an {{Extended Gutzwiller Approximation}} for t--{{J Model}}: {{Effect}} of {{Double-Occupancy Exclusion}}},
  shorttitle = {Superconductivity and {{Antiferromagnetism}} in an {{Extended Gutzwiller Approximation}} for t--{{J Model}}},
  author = {Ogata, Masao and Himeda, Akihiro},
  year = {2003},
  month = feb,
  journal = {Journal of the Physical Society of Japan},
  volume = {72},
  number = {2},
  pages = {374--391},
  publisher = {The Physical Society of Japan},
  issn = {0031-9015},
  doi = {10.1143/JPSJ.72.374},
  urldate = {2025-09-09},
}

@article{2019Hwang,
    author = {Li, Danfeng and Lee, Kyuho and Wang, Bai Yang and Osada, Motoki and Crossley, Samuel and Lee, Hye Ryoung and Cui, Yi and Hikita, Yasuyuki and Hwang, Harold Y.},
    da = {2019/08/01},
    doi = {10.1038/s41586-019-1496-5},
    id = {Li2019},
    isbn = {1476-4687},
    journal = {Nature},
    number = {7771},
    pages = {624--627},
    title = {Superconductivity in an infinite-layer nickelate},
    ty = {JOUR},
    volume = {572},
    year = {2019},
}

@article{2024Zhaoa,
  title = {Superconductivity in Pressurized Trilayer {{La4Ni3O10-}}{{$\delta$}} Single Crystals},
  author = {Zhu, Yinghao and Peng, Di and Zhang, Enkang and Pan, Bingying and Chen, Xu and Chen, Lixing and Ren, Huifen and Liu, Feiyang and Hao, Yiqing and Li, Nana and Xing, Zhenfang and Lan, Fujun and Han, Jiyuan and Wang, Junjie and Jia, Donghan and Wo, Hongliang and Gu, Yiqing and Gu, Yimeng and Ji, Li and Wang, Wenbin and Gou, Huiyang and Shen, Yao and Ying, Tianping and Chen, Xiaolong and Yang, Wenge and Cao, Huibo and Zheng, Changlin and Zeng, Qiaoshi and Guo, Jian-gang and Zhao, Jun},
  year = 2024,
  month = jul,
  journal = {Nature},
  volume = {631},
  number = {8021},
  pages = {531--536},
  publisher = {Nature Publishing Group},
  issn = {1476-4687},
  doi = {10.1038/s41586-024-07553-3},
  urldate = {2025-02-24},
  copyright = {2024 The Author(s), under exclusive licence to Springer Nature Limited},
  langid = {english},
}

@article{2025Qi,
  title = {Superconductivity in Trilayer Nickelate ${\mathrm{La}}_{4}{\mathrm{Ni}}_{3}{\mathrm{O}}_{10}$ under Pressure},
  author = {Zhang, Mingxin and Pei, Cuiying and Peng, Di and Du, Xian and Hu, Weixiong and Cao, Yantao and Wang, Qi and Wu, Juefei and Li, Yidian and Liu, Huanyu and Wen, Chenhaoping and Song, Jing and Zhao, Yi and Li, Changhua and Cao, Weizheng and Zhu, Shihao and Zhang, Qing and Yu, Na and Cheng, Peihong and Zhang, Lili and Li, Zhiwei and Zhao, Jinkui and Chen, Yulin and Jin, Changqing and Guo, Hanjie and Wu, Congjun and Yang, Fan and Zeng, Qiaoshi and Yan, Shichao and Yang, Lexian and Qi, Yanpeng},
  journal = {Physical Review X},
  volume = {15},
  issue = {2},
  pages = {021005},
  numpages = {11},
  year = {2025},
  month = {Apr},
  publisher = {American Physical Society},
  doi = {10.1103/PhysRevX.15.021005},
  url = {https://link.aps.org/doi/10.1103/PhysRevX.15.021005}
}

@ARTICLE{1973KK,
       author = {{Kugel'}, K.~I. and {Khomski{\v{i}}}, D.~I.},
        title = "{Crystal structure and magnetic properties of substances with orbital degeneracy}",
      journal = {Soviet Journal of Experimental and Theoretical Physics},
         year = 1973,
        month = oct,
       volume = {37},
        pages = {725},
       adsurl = {https://ui.adsabs.harvard.edu/abs/1973JETP...37..725K}
}

@article{2016Xue,
title = {Nodeless pairing in superconducting copper-oxide monolayer films on Bi2Sr2CaCu2O8+δ},
journal = {Science Bulletin},
volume = {61},
number = {16},
pages = {1239-1247},
year = {2016},
issn = {2095-9273},
doi = {https://doi.org/10.1007/s11434-016-1145-4},
url = {https://www.sciencedirect.com/science/article/pii/S2095927316300494},
author = {Yong Zhong and Yang Wang and Sha Han and Yan-Feng Lv and Wen-Lin Wang and Ding Zhang and Hao Ding and Yi-Min Zhang and Lili Wang and Ke He and Ruidan Zhong and John A. Schneeloch and Gen-Da Gu and Can-Li Song and Xu-Cun Ma and Qi-Kun Xue},
}

@article{2018Jiang,
  title = {Nodeless High-${T}_{c}$ Superconductivity in the Highly Overdoped ${\mathrm{CuO}}_{2}$ Monolayer},
  author = {Jiang, Kun and Wu, Xianxin and Hu, Jiangping and Wang, Ziqiang},
  journal = {Physical Review Letters},
  volume = {121},
  issue = {22},
  pages = {227002},
  numpages = {6},
  year = {2018},
  month = {Nov},
  publisher = {American Physical Society},
  doi = {10.1103/PhysRevLett.121.227002},
  url = {https://link.aps.org/doi/10.1103/PhysRevLett.121.227002}
}

@article{2008Khaliullin,
  title = {Orbital Order and Possible Superconductivity in ${\mathrm{LaNiO}}_{3}/{\mathrm{LaMO}}_{3}$ Superlattices},
  author = {Chaloupka, Ji\ifmmode \check{r}\else \v{r}\fi{}\'{\i} and Khaliullin, Giniyat},
  journal = {Physical Review Letters},
  volume = {100},
  issue = {1},
  pages = {016404},
  numpages = {4},
  year = {2008},
  month = {Jan},
  publisher = {American Physical Society},
  doi = {10.1103/PhysRevLett.100.016404},
  url = {https://link.aps.org/doi/10.1103/PhysRevLett.100.016404}
}

@article{2019Jin,
author = {W. M. Li  and J. F. Zhao  and L. P. Cao  and Z. Hu  and Q. Z. Huang  and X. C. Wang  and Y. Liu  and G. Q. Zhao  and J. Zhang  and Q. Q. Liu  and R. Z. Yu  and Y. W. Long  and H. Wu  and H. J. Lin  and C. T. Chen  and Z. Li  and Z. Z. Gong  and Z. Guguchia  and J. S. Kim  and G. R. Stewart  and Y. J. Uemura  and S. Uchida  and C. Q. Jin },
title = {Superconductivity in a unique type of copper oxide},
journal = {Proceedings of the National Academy of Sciences},
volume = {116},
number = {25},
pages = {12156-12160},
year = {2019},
doi = {10.1073/pnas.1900908116},
URL = {https://www.pnas.org/doi/abs/10.1073/pnas.1900908116},
}

@article{2025Jiang,
  title = {Electronic Structure and Disorder Effect of {{La3Ni2O7}} Superconductor},
  author = {Wang, Yuxin and Zhang, Yi and Jiang, Kun},
  year = {2025},
  month = apr,
  journal = {Chinese Physics B},
  volume = {34},
  number = {4},
  pages = {047105},
  publisher = {{Chinese Physical Society and IOP Publishing Ltd}},
  issn = {1674-1056},
  doi = {10.1088/1674-1056/adbacc},
  urldate = {2025-05-17},
  langid = {english}
}

@article{1988Shibab,
  title = {Variational {{Monte-Carlo Studies}} of {{Superconductivity}} in {{Strongly Correlated Electron Systems}}},
  author = {Yokoyama, Hisatoshi and Shiba, Hiroyuki},
  year = {1988},
  month = jul,
  journal = {Journal of the Physical Society of Japan},
  volume = {57},
  number = {7},
  pages = {2482--2493},
  publisher = {The Physical Society of Japan},
  issn = {0031-9015},
  doi = {10.1143/JPSJ.57.2482},
  urldate = {2025-09-09}
}

@article{2004Lee,
  title = {Absence of the Coexistence of Superconductivity and Antiferromagnetism in the Hole-Doped Two-Dimensional Extended \$t{\textbackslash}text\{{\textbackslash}ensuremath\{-\}\}{{J}}\$ Model},
  author = {Shih, C. T. and Chen, Y. C. and Chou, C. P. and Lee, T. K.},
  year = {2004},
  month = dec,
  journal = {Physical Review B},
  volume = {70},
  number = {22},
  pages = {220502},
  publisher = {American Physical Society},
  doi = {10.1103/PhysRevB.70.220502},
  urldate = {2025-09-09},
}

@article{2023Wangc,
  title = {Signatures of Superconductivity near 80 {{K}} in a Nickelate under High Pressure},
  author = {Sun, Hualei and Huo, Mengwu and Hu, Xunwu and Li, Jingyuan and Liu, Zengjia and Han, Yifeng and Tang, Lingyun and Mao, Zhongquan and Yang, Pengtao and Wang, Bosen and Cheng, Jinguang and Yao, Dao-Xin and Zhang, Guang-Ming and Wang, Meng},
  year = 2023,
  month = sep,
  journal = {Nature},
  volume = {621},
  number = {7979},
  pages = {493--498},
  publisher = {Nature Publishing Group},
  issn = {1476-4687},
  doi = {10.1038/s41586-023-06408-7},
  urldate = {2024-11-17},
  langid = {english},
}

@article{2025Wangi,
  title = {Evolution of the Superconductivity in Pressurized {{La3-xSmxNi2O7}}},
  author = {Zhong, Qingyi and Chen, Junfeng and Qiu, Zhengyang and Li, Jingyuan and Huang, Xing and Ma, Peiyue and Huo, Mengwu and Dong, Hongliang and Sun, Hualei and Wang, Meng},
  year = 2025,
  month = oct,
  number = {arXiv:2510.13342},
  eprint = {2510.13342},
  primaryclass = {cond-mat},
  publisher = {arXiv},
  urldate = {2025-10-16},
  archiveprefix = {arXiv},
  journal={},
}

@article{2025Wangg,
  title = {Interlayer Coupling Enhanced Superconductivity near 100 {{K}} in {{La}}\$\_\textbraceleft 3-x\textbraceright\${{Nd}}\$\_x\${{Ni}}\$\_2\${{O}}\$\_7\$},
  author = {Qiu, Zhengyang and Chen, Junfeng and Semenok, Dmitrii V. and Zhong, Qingyi and Zhou, Di and Li, Jingyuan and Ma, Peiyue and Huang, Xing and Huo, Mengwu and Xie, Tao and Chen, Xiang and Mao, Ho-kwang and Struzhkin, Viktor and Sun, Hualei and Wang, Meng},
  year = 2025,
  month = oct,
  number = {arXiv:2510.12359},
  eprint = {2510.12359},
  primaryclass = {cond-mat},
  publisher = {arXiv},
  journal = {}, 
  urldate = {2025-10-15},
  archiveprefix = {arXiv}
}

@article{2025Wang,
  title = {Identification of Superconductivity in Bilayer Nickelate {{La3Ni2O7}} under High Pressure up to 100 {{GPa}}},
  author = {Li, Jingyuan and Peng, Di and Ma, Peiyue and Zhang, Hengyuan and Xing, Zhenfang and Huang, Xing and Huang, Chaoxin and Huo, Mengwu and Hu, Deyuan and Dong, Zixian and Chen, Xiang and Xie, Tao and Dong, Hongliang and Sun, Hualei and Zeng, Qiaoshi and Mao, Ho-kwang and Wang, Meng},
  year = 2025,
  month = oct,
  journal = {National Science Review},
  volume = {12},
  number = {10},
  pages = {nwaf220},
  issn = {2095-5138},
  doi = {10.1093/nsr/nwaf220},
  urldate = {2025-11-07},
}

@article{chengjg_crystal,
  title = {Bulk High-Temperature Superconductivity in Pressurized Tetragonal {{La2PrNi2O7}}},
  author = {Wang, Ningning and Wang, Gang and Shen, Xiaoling and Hou, Jun and Luo, Jun and Ma, Xiaoping and Yang, Huaixin and Shi, Lifen and Dou, Jie and Feng, Jie and Yang, Jie and Shi, Yunqing and Ren, Zhian and Ma, Hanming and Yang, Pengtao and Liu, Ziyi and Liu, Yue and Zhang, Hua and Dong, Xiaoli and Wang, Yuxin and Jiang, Kun and Hu, Jiangping and Nagasaki, Shoko and Kitagawa, Kentaro and Calder, Stuart and Yan, Jiaqiang and Sun, Jianping and Wang, Bosen and Zhou, Rui and Uwatoko, Yoshiya and Cheng, Jinguang},
  year = 2024,
  month = oct,
  journal = {Nature},
  volume = {634},
  number = {8034},
  pages = {579--584},
  publisher = {Nature Publishing Group},
  issn = {1476-4687},
  doi = {10.1038/s41586-024-07996-8},
  urldate = {2024-11-20},
  copyright = {2024 The Author(s), under exclusive licence to Springer Nature Limited},
  langid = {english}
}

@article{chengjg_poly,
  title = {Pressure-Induced Superconductivity in Polycrystalline {{La}}{$_{3}$}{{Ni}}{$_{2}$}{{O}}{\textsubscript{7-{$\delta$}}}},
  author = {Wang, G. and Wang, N. N. and Shen, X. L. and Hou, J. and Ma, L. and Shi, L. F. and Ren, Z. A. and Gu, Y. D. and Ma, H. M. and Yang, P. T. and Liu, Z. Y. and Guo, H. Z. and Sun, J. P. and Zhang, G. M. and Calder, S. and Yan, J.-Q. and Wang, B. S. and Uwatoko, Y. and Cheng, J.-G.},
  year = 2024,
  month = mar,
  journal = {Physical Review X},
  volume = {14},
  number = {1},
  pages = {011040},
  publisher = {American Physical Society},
  doi = {10.1103/PhysRevX.14.011040}
}

@article{2024Yuan,
  title = {High-Temperature Superconductivity with Zero Resistance and Strange-Metal Behaviour in {{La3Ni2O7}}-{$\delta$}},
  author = {Zhang, Yanan and Su, Dajun and Huang, Yanen and Shan, Zhaoyang and Sun, Hualei and Huo, Mengwu and Ye, Kaixin and Zhang, Jiawen and Yang, Zihan and Xu, Yongkang and Su, Yi and Li, Rui and Smidman, Michael and Wang, Meng and Jiao, Lin and Yuan, Huiqiu},
  year = 2024,
  month = aug,
  journal = {Nature Physics},
  volume = {20},
  number = {8},
  pages = {1269--1273},
  publisher = {Nature Publishing Group},
  issn = {1745-2481},
  doi = {10.1038/s41567-024-02515-y},
  urldate = {2025-02-26},
  copyright = {2024 The Author(s), under exclusive licence to Springer Nature Limited},
  langid = {english},
}

@article{2025Maoa,
  title = {Direct {{Observation}} of D-{{Wave Superconducting Gap Symmetry}} in {{Pressurized La3Ni2O7-delta Single Crystals}}},
  author = {Cao, Zi-Yu and Peng, Di and Choi, Seokmin and Lan, Fujun and Yu, Lan and Zhang, Enkang and Xing, Zhenfang and Liu, Yuxin and Zhang, Feiyang and Luo, Tao and Chen, Lixing and Hong, Vuong Thi Anh and Paek, Seung-Yeop and Jang, Harim and Xie, Jinghong and Liu, Huayu and Lou, Hongbo and Zeng, Zhidan and Ding, Yang and Zhao, Jun and Liu, Cailong and Park, Tuson and Zeng, Qiaoshi and Mao, Ho-kwang},
  year = 2025,
  month = sep,
  number = {arXiv:2509.12606},
  eprint = {2509.12606},
  primaryclass = {cond-mat},
  publisher = {arXiv},
  journal = {},
  urldate = {2025-09-17},
  archiveprefix = {arXiv},
}

@article{2025Hwangc,
  title = {Signatures of Ambient Pressure Superconductivity in Thin Film {{La3Ni2O7}}},
  author = {Ko, Eun Kyo and Yu, Yijun and Liu, Yidi and Bhatt, Lopa and Li, Jiarui and Thampy, Vivek and Kuo, Cheng-Tai and Wang, Bai Yang and Lee, Yonghun and Lee, Kyuho and Lee, Jun-Sik and Goodge, Berit H. and Muller, David A. and Hwang, Harold Y.},
  year = 2025,
  month = feb,
  journal = {Nature},
  volume = {638},
  number = {8052},
  pages = {935--940},
  issn = {1476-4687},
  doi = {10.1038/s41586-024-08525-3},
}

@article{2025Hwang,
  title = {Superconductivity and Normal-State Transport in Compressively Strained {{La2PrNi2O7}} Thin Films},
  author = {Liu, Yidi and Ko, Eun Kyo and Tarn, Yaoju and Bhatt, Lopa and Li, Jiarui and Thampy, Vivek and Goodge, Berit H. and Muller, David A. and Raghu, Srinivas and Yu, Yijun and Hwang, Harold Y.},
  year = 2025,
  month = may,
  journal = {Nature Materials},
  pages = {1--7},
  publisher = {Nature Publishing Group},
  issn = {1476-4660},
  doi = {10.1038/s41563-025-02258-y},
  urldate = {2025-06-11},
  copyright = {2025 The Author(s), under exclusive licence to Springer Nature Limited},
  langid = {english},
}

@article{2025Hwanga,
  title = {Fermi-Liquid Transport beyond the Upper Critical Field in Superconducting {{La}}\$\_2\${{PrNi}}\$\_2\${{O}}\$\_7\$ Thin Films},
  author = {Hsu, Yu-Te and Liu, Yidi and Kohama, Yoshimitsu and Kotte, Tommy and Sharma, Vikash and Tarn, Yaoju and Yu, Yijun and Hwang, Harold Y.},
  year = 2025,
  month = may,
  number = {arXiv:2505.19011},
  eprint = {2505.19011},
  primaryclass = {cond-mat},
  publisher = {arXiv},
  journal = {},
  urldate = {2025-05-28},
  archiveprefix = {arXiv},
}

@article{2025Hwangb,
  title = {Reducing the Strain Required for Ambient-Pressure Superconductivity in Bilayer Nickelates},
  author = {Tarn, Yaoju and Liu, Yidi and Theuss, Florian and Li, Jiarui and Wang, Bai Yang and Wang, Jiayue and Thampy, Vivek and Shen, Zhi-Xun and Yu, Yijun and Hwang, Harold Y.},
  year = 2025,
  month = oct,
  number = {arXiv:2510.27613},
  eprint = {2510.27613},
  primaryclass = {cond-mat},
  publisher = {arXiv},
  journal = {},
  urldate = {2025-11-07},
  archiveprefix = {arXiv},
}

@article{2025Shen,
  title = {Electronic Structure of Compressively Strained Thin Film {{La}}\$\_2\${{PrNi}}\$\_2\${{O}}\$\_7\$},
  author = {Wang, Bai Yang and Zhong, Yong and Abadi, Sebastien and Liu, Yidi and Yu, Yijun and Zhang, Xiaoliang and Wu, Yi-Ming and Wang, Ruohan and Li, Jiarui and Tarn, Yaoju and Ko, Eun Kyo and Thampy, Vivek and Hashimoto, Makoto and Lu, Donghui and Lee, Young S. and Devereaux, Thomas P. and Jia, Chunjing and Hwang, Harold Y. and Shen, Zhi-Xun},
  year = 2025,
  month = apr,
  number = {arXiv:2504.16372},
  eprint = {2504.16372},
  primaryclass = {cond-mat},
  publisher = {arXiv},
  journal = {},
  urldate = {2025-04-24},
  archiveprefix = {arXiv},
}

@article{2025Chenb,
  title = {Ambient-Pressure Superconductivity Onset above 40 {{K}} in ({{La}},{{Pr}}){{3Ni2O7}} Films},
  author = {Zhou, Guangdi and Lv, Wei and Wang, Heng and Nie, Zihao and Chen, Yaqi and Li, Yueying and Huang, Haoliang and Chen, Wei-Qiang and Sun, Yu-Jie and Xue, Qi-Kun and Chen, Zhuoyu},
  year = 2025,
  month = apr,
  journal = {Nature},
  volume = {640},
  number = {8059},
  pages = {641--646},
  issn = {0028-0836, 1476-4687},
  doi = {10.1038/s41586-025-08755-z},
  urldate = {2025-06-30},
  langid = {english},
}

@article{2025Xuea,
  title = {Angle-Resolved Photoemission Spectroscopy of Superconducting ({{La}},{{Pr}}){{3Ni2O7}}/{{SrLaAlO4}} Heterostructures},
  author = {Li, Peng and Zhou, Guangdi and Lv, Wei and Li, Yueying and Yue, Changming and Huang, Haoliang and Xu, Lizhi and Shen, Jianchang and Miao, Yu and Song, Wenhua and Nie, Zihao and Chen, Yaqi and Wang, Heng and Chen, Weiqiang and Huang, Yaobo and Chen, Zhen-Hua and Qian, Tian and Lin, Junhao and He, Junfeng and Sun, Yu-Jie and Chen, Zhuoyu and Xue, Qi-Kun},
  year = 2025,
  month = may,
  journal = {National Science Review},
  pages = {nwaf205},
  issn = {2095-5138},
  doi = {10.1093/nsr/nwaf205},
  urldate = {2025-06-01},
}

@article{2024Chenc,
  title = {Visualization of Oxygen Vacancies and Self-Doped Ligand Holes in {{La3Ni2O7}}-{$\delta$}},
  author = {Dong, Zehao and Huo, Mengwu and Li, Jie and Li, Jingyuan and Li, Pengcheng and Sun, Hualei and Gu, Lin and Lu, Yi and Wang, Meng and Wang, Yayu and Chen, Zhen},
  year = 2024,
  month = jun,
  journal = {Nature},
  volume = {630},
  number = {8018},
  pages = {847--852},
  issn = {0028-0836, 1476-4687},
  doi = {10.1038/s41586-024-07482-1},
  urldate = {2025-02-26},
  langid = {english},
}

@article{2025Wangd,
  title = {Interstitial Oxygen Order and Its Competition with Superconductivity in {{La2PrNi2O7}}+{$\delta$}},
  author = {Dong, Zehao and Wang, Gang and Wang, Ningning and Dong, Wen-Han and Gu, Lin and Xu, Yong and Cheng, Jinguang and Chen, Zhen and Wang, Yayu},
  year = 2025,
  month = dec,
  journal = {Nature Materials},
  volume = {24},
  number = {12},
  pages = {1927--1934},
  issn = {1476-1122, 1476-4660},
  doi = {10.1038/s41563-025-02351-2},
  urldate = {2026-03-10},
  langid = {english},
}

@article{2024Feng,
  title = {Electronic and Magnetic Excitations in {{La3Ni2O7}}},
  author = {Chen, Xiaoyang and Choi, Jaewon and Jiang, Zhicheng and Mei, Jiong and Jiang, Kun and Li, Jie and Agrestini, Stefano and {Garcia-Fernandez}, Mirian and Sun, Hualei and Huang, Xing and Shen, Dawei and Wang, Meng and Hu, Jiangping and Lu, Yi and Zhou, Ke-Jin and Feng, Donglai},
  year = 2024,
  month = nov,
  journal = {Nature Communications},
  volume = {15},
  number = {1},
  pages = {9597},
  publisher = {Nature Publishing Group},
  issn = {2041-1723},
  doi = {10.1038/s41467-024-53863-5},
  urldate = {2025-03-09},
  copyright = {2024 The Author(s)},
  langid = {english},
}

@article{2024Mathon,
  title = {Local Electronic Properties of {{La3Ni2O7}} under Pressure},
  author = {Mijit, Emin and Ma, Peiyue and Sahle, Christoph J. and Rosa, Angelika D. and Hu, Zhiwei and Angelis, Francesco De and Lopez, Alberto and Amatori, Simone and Tchoudinov, Georghii and Joly, Yves and Irifune, Tetsuo and Rodrigues, Joao Elias F. S. and Garbarino, Gaston and Parra, Samuel Gallego and Wang, Meng and Yu, Runze and Mathon, Olivier},
  year = 2024,
  month = dec,
  number = {arXiv:2412.08269},
  eprint = {2412.08269},
  primaryclass = {cond-mat},
  publisher = {arXiv},
  journal = {},
  urldate = {2025-06-12},
  archiveprefix = {arXiv},
  langid = {english},
}

@article{2025Chenc,
  title = {Pressure Induced Superconductivity in Hybrid {{Ruddlesden}}-{{Popper La5Ni3O11}} Single Crystals},
  author = {Shi, Mengzhu and Peng, Di and Fan, Kaibao and Xing, Zhenfang and Yang, Shaohua and Wang, Yuzhu and Li, Houpu and Wu, Rongqi and Du, Mei and Ge, Binghui and Zeng, Zhidan and Zeng, Qiaoshi and Ying, Jianjun and Wu, Tao and Chen, Xianhui},
  year = 2025,
  month = nov,
  journal = {Nature Physics},
  volume = {21},
  number = {11},
  pages = {1780--1786},
  publisher = {Nature Publishing Group},
  issn = {1745-2481},
  doi = {10.1038/s41567-025-03023-3},
  urldate = {2026-01-01},
  copyright = {2025 The Author(s), under exclusive licence to Springer Nature Limited},
  langid = {english}
}

@article{2025Chene,
  title = {Recent Progress in Nickelate Superconductors},
  author = {Wang, Yuxin and Jiang, Kun and Ying, Jianjun and Wu, Tao and Cheng, Jinguang and Hu, Jiangping and Chen, Xianhui},
  year = 2025,
  month = sep,
  journal = {National Science Review},
  volume = {12},
  number = {10},
  pages = {nwaf373},
  issn = {2095-5138, 2053-714X},
  doi = {10.1093/nsr/nwaf373},
  urldate = {2025-11-05},
  copyright = {https://creativecommons.org/licenses/by/4.0/},
  langid = {english},
}

@article{2025He,
  title = {Anomalous Energy Gap in Superconducting {{La}}\$\_\textbraceleft 2.85\textbraceright\${{Pr}}\$\_\textbraceleft 0.15\textbraceright\${{Ni}}\$\_2\${{O}}\$\_7\$/{{SrLaAlO}}\$\_4\$ Heterostructures},
  author = {Shen, Jianchang and Miao, Yu and Ou, Zhipeng and Zhou, Guangdi and Chen, Yaqi and Luan, Runqing and Sun, Hongxu and Feng, Zikun and Yong, Xinru and Li, Peng and Li, Yueying and Xu, Lizhi and Lv, Wei and Nie, Zihao and Wang, Heng and Huang, Haoliang and Sun, Yu-Jie and Xue, Qi-Kun and Chen, Zhuoyu and He, Junfeng},
  year = 2025,
  month = feb,
  number = {arXiv:2502.17831},
  eprint = {2502.17831},
  primaryclass = {cond-mat},
  publisher = {arXiv},
  journal = {},
  urldate = {2025-04-23},
  archiveprefix = {arXiv},
}

@article{2025Nie,
  title = {Superconductivity in {{Sr-doped La3Ni2O7}} Thin Films},
  author = {Hao, Bo and Wang, Maosen and Sun, Wenjie and Yang, Yang and Mao, Zhangwen and Yan, Shengjun and Sun, Haoying and Zhang, Hongyi and Han, Lu and Gu, Zhengbin and Zhou, Jian and Ji, Dianxiang and Nie, Yuefeng},
  year = 2025,
  month = nov,
  journal = {Nature Materials},
  volume = {24},
  number = {11},
  pages = {1756--1762},
  issn = {1476-1122, 1476-4660},
  doi = {10.1038/s41563-025-02327-2},
  urldate = {2025-12-06},
  langid = {english},
}

@article{2025Niea,
      title={Observation of superconductivity-induced leading-edge gap in Sr-doped $\mathrm{La}_{3}\mathrm{Ni}_{2}\mathrm{O}_{7}$ thin films},
      author={Wenjie Sun and Zhicheng Jiang and Bo Hao and Shengjun Yan and Hongyi Zhang and Maosen Wang and Yang Yang and Haoying Sun and Zhengtai Liu and Dianxiang Ji and Zhengbin Gu and Jian Zhou and Dawei Shen and Donglai Feng and Yuefeng Nie},
      year={2025},
      eprint={2507.07409},
      archivePrefix={arXiv},
      primaryClass={cond-mat.supr-con},
      journal = {},
}

@article{2025Zhanga,
	author = {Li, Feiyu and Xing, Zhenfang and Peng, Di and Dou, Jie and Guo, Ning and Ma, Liang and Zhang, Yulin and Wang, Lingzhen and Luo, Jun and Yang, Jie and Zhang, Jian and Chang, Tieyan and Chen, Yu-Sheng and Cai, Weizhao and Cheng, Jinguang and Wang, Yuzhu and Liu, Yuxin and Luo, Tao and Hirao, Naohisa and Matsuoka, Takahiro and Kadobayashi, Hirokazu and Zeng, Zhidan and Zheng, Qiang and Zhou, Rui and Zeng, Qiaoshi and Tao, Xutang and Zhang, Junjie},
	date = {2026/01/01},
	doi = {10.1038/s41586-025-09954-4},
	id = {Li2026},
	isbn = {1476-4687},
	journal = {Nature},
	number = {8098},
	pages = {871--878},
	title = {Bulk superconductivity up to 96 K in pressurized nickelate single crystals},
	url = {https://doi.org/10.1038/s41586-025-09954-4},
	volume = {649},
	year = {2026}
}

@article{2025Tsukazaki,
  title = {Strain-Tuning for Superconductivity in {{La3Ni2O7}} Thin Films},
  author = {Osada, Motoki and Terakura, Chieko and Kikkawa, Akiko and Nakajima, Masamichi and Chen, Hsiao-Yi and Nomura, Yusuke and Tokura, Yoshinori and Tsukazaki, Atsushi},
  year = 2025,
  month = jun,
  journal = {Communications Physics},
  volume = {8},
  number = {1},
  pages = {251},
  publisher = {Nature Publishing Group},
  issn = {2399-3650},
  doi = {10.1038/s42005-025-02154-6},
  urldate = {2025-11-07},
  copyright = {2025 The Author(s)},
  langid = {english},
}

@article{2023Yao,
  title = {Bilayer Two-Orbital Model of $\mathrm{L}{\mathrm{a}}_{3}\mathrm{N}{\mathrm{i}}_{2}{\mathrm{O}}_{7}$ under Pressure},
  author = {Luo, Zhihui and Hu, Xunwu and Wang, Meng and W{\'u}, W{\'e}i and Yao, Dao-Xin},
  year = 2023,
  month = sep,
  journal = {Physical Review Letters},
  volume = {131},
  number = {12},
  pages = {126001},
  publisher = {American Physical Society},
  doi = {10.1103/PhysRevLett.131.126001},
  urldate = {2025-02-24},
}

@article{2023Yangb,
  title = {\$\textbraceleft s\textbraceright\textasciicircum\textbraceleft\textbackslash ifmmode\textbackslash pm\textbackslash else\textbackslash textpm\textbackslash fi\textbraceleft\textbraceright\textbraceright\$-{{Wave Pairing}} and the {{Destructive Role}} of {{Apical-Oxygen Deficiencies}} in \$\textbraceleft\textbackslash mathrm\textbraceleft{{La}}\textbraceright\textbraceright\_\textbraceleft 3\textbraceright\textbraceleft\textbackslash mathrm\textbraceleft{{Ni}}\textbraceright\textbraceright\_\textbraceleft 2\textbraceright\textbraceleft\textbackslash mathrm\textbraceleft{{O}}\textbraceright\textbraceright\_\textbraceleft 7\textbraceright\$ under {{Pressure}}},
  author = {Liu, Yu-Bo and Mei, Jia-Wei and Ye, Fei and Chen, Wei-Qiang and Yang, Fan},
  year = 2023,
  month = dec,
  journal = {Physical Review Letters},
  volume = {131},
  number = {23},
  pages = {236002},
  publisher = {American Physical Society},
  doi = {10.1103/PhysRevLett.131.236002},
  urldate = {2025-02-24},
}

@article{2024Wua,
  title = {Interlayer-{{Coupling-Driven High-Temperature Superconductivity}} in \$\textbraceleft\textbackslash mathrm\textbraceleft{{La}}\textbraceright\textbraceright\_\textbraceleft 3\textbraceright\textbraceleft\textbackslash mathrm\textbraceleft{{Ni}}\textbraceright\textbraceright\_\textbraceleft 2\textbraceright\textbraceleft\textbackslash mathrm\textbraceleft{{O}}\textbraceright\textbraceright\_\textbraceleft 7\textbraceright\$ under {{Pressure}}},
  author = {Lu, Chen and Pan, Zhiming and Yang, Fan and Wu, Congjun},
  year = 2024,
  month = apr,
  journal = {Physical Review Letters},
  volume = {132},
  number = {14},
  pages = {146002},
  publisher = {American Physical Society},
  doi = {10.1103/PhysRevLett.132.146002},
  urldate = {2025-02-24},
}

@article{2025Chen,
  title = {Correlated Electronic Structures and Unconventional Superconductivity in Bilayer Nickelate Heterostructures},
  author = {Yue, Changming and Miao, Jian-Jian and Huang, Haoliang and Hua, Yichen and Li, Peng and Li, Yueying and Zhou, Guangdi and Lv, Wei and Yang, Qishuo and Yang, Fan and Sun, Hongyi and Sun, Yu-Jie and Lin, Junhao and Xue, Qi-Kun and Chen, Zhuoyu and Chen, Wei-Qiang},
  year = 2025,
  month = oct,
  journal = {National Science Review},
  volume = {12},
  number = {10},
  pages = {nwaf253},
  issn = {2095-5138},
  doi = {10.1093/nsr/nwaf253},
  urldate = {2026-04-01},
}

@article{2023Si,
  title = {Electron correlations and superconductivity in ${\mathrm{La}}_{3}{\mathrm{Ni}}_{2}{\mathrm{O}}_{7}$ under pressure tuning},
  author = {Liao, Zhiguang and Chen, Lei and Duan, Guijing and Wang, Yiming and Liu, Changle and Yu, Rong and Si, Qimiao},
  year = 2023,
  month = dec,
  journal = {Physical Review B},
  volume = {108},
  number = {21},
  pages = {214522},
  publisher = {American Physical Society},
  doi = {10.1103/PhysRevB.108.214522},
  urldate = {2025-02-24},
}

@article{2026Raghu,
      title={Superconductivity and magnetism in bilayer nickelates: itinerant perspective},
      author={Yi-Ming Wu and Tobias Helbig and Salahudin V. Smailagić and Hao-Xin Wang and Yijun Yu and Harold Y. Hwang and Srinivas Raghu},
      year={2026},
      eprint={2602.20288},
      archivePrefix={arXiv},
      primaryClass={cond-mat.supr-con},
      journal={},
}

@article{2023Eremin,
  title = {Electronic correlations and superconducting instability in ${\mathrm{La}}_{3}{\mathrm{Ni}}_{2}{\mathrm{O}}_{7}$ under high pressure},
  author = {Lechermann, Frank and Gondolf, Jannik and B{\"o}tzel, Steffen and Eremin, Ilya M.},
  year = 2023,
  month = nov,
  journal = {Physical Review B},
  volume = {108},
  number = {20},
  pages = {L201121},
  publisher = {American Physical Society},
  doi = {10.1103/PhysRevB.108.L201121},
  urldate = {2025-02-24},
}

@article{2025Hug,
  title = {Effective model and pairing tendency in the bilayer Ni-based superconductor ${\mathrm{La}}_{3}{\mathrm{Ni}}_{2}{\mathrm{O}}_{7}$},
  author = {Gu, Yuhao and Le, Congcong and Yang, Zhesen and Wu, Xianxin and Hu, Jiangping},
  journal = {Physical Review B},
  volume = {111},
  issue = {17},
  pages = {174506},
  numpages = {7},
  year = {2025},
  month = {May},
  publisher = {American Physical Society},
  doi = {10.1103/PhysRevB.111.174506},
  url = {https://link.aps.org/doi/10.1103/PhysRevB.111.174506}
}

@article{2025Hua,
  title = {Cooperation between Electron-Phonon Coupling and Electronic Interaction in Bilayer Nickelates ${\mathrm{La}}_{3}{\mathrm{Ni}}_{2}{\mathrm{O}}_{7}$},
  author = {Zhan, Jun and Gu, Yuhao and Wu, Xianxin and Hu, Jiangping},
  journal = {Physical Review Letters},
  volume = {134},
  issue = {13},
  pages = {136002},
  numpages = {7},
  year = {2025},
  month = {Mar},
  publisher = {American Physical Society},
  doi = {10.1103/PhysRevLett.134.136002},
  url = {https://link.aps.org/doi/10.1103/PhysRevLett.134.136002}
}

@article{2024Hub-dft,
  title = {Electronic and Magnetic Structures of Bilayer \$\textbraceleft\textbackslash text\textbraceleft{{La}}\textbraceright\textbraceright\_\textbraceleft 3\textbraceright\textbraceleft\textbackslash text\textbraceleft{{Ni}}\textbraceright\textbraceright\_\textbraceleft 2\textbraceright\textbraceleft\textbackslash text\textbraceleft{{O}}\textbraceright\textbraceright\_\textbraceleft 7\textbraceright\$ at Ambient Pressure},
  author = {Wang, Yuxin and Jiang, Kun and Wang, Ziqiang and Zhang, Fu-Chun and Hu, Jiangping},
  year = 2024,
  month = nov,
  journal = {Physical Review B},
  volume = {110},
  number = {20},
  pages = {205122},
  publisher = {American Physical Society},
  doi = {10.1103/PhysRevB.110.205122},
  urldate = {2025-02-24},
}

@article{2025Hua-dft,
  title = {The {{Mottness}} and the {{Anderson}} Localization in Bilayer Nickelate {{La}}\$\_3\${{Ni}}\$\_2\${{O}}\$\_\textbraceleft 7-{$\delta$}\textbraceright\$},
  author = {Wang, Yuxin and Chen, Ziyan and Zhang, Yi and Jiang, Kun and Hu, Jiangping},
  year = 2025,
  month = jan,
  number = {arXiv:2501.08536},
  eprint = {2501.08536},
  primaryclass = {cond-mat},
  publisher = {arXiv},
  journal={},
  urldate = {2025-04-28},
}

@article{2025Jiang-dft,
  title = {Electronic Structure and Disorder Effect of {{La3Ni2O7}} Superconductor},
  author = {Wang, Yuxin and Zhang, Yi and Jiang, Kun},
  year = 2025,
  month = apr,
  journal = {Chinese Physics B},
  volume = {34},
  number = {4},
  pages = {047105},
  publisher = {{Chinese Physical Society and IOP Publishing Ltd}},
  issn = {1674-1056},
  doi = {10.1088/1674-1056/adbacc},
  urldate = {2025-05-17},
}

@article{2023Wangd,
  title = {Possible ${s}_{\ifmmode\pm\else\textpm\fi{}}$-wave superconductivity in ${\mathrm{La}}_{3}{\mathrm{Ni}}_{2}{\mathrm{O}}_{7}$},
  author = {Yang, Qing-Geng and Wang, Da and Wang, Qiang-Hua},
  year = 2023,
  month = oct,
  journal = {Physical Review B},
  volume = {108},
  number = {14},
  pages = {L140505},
  publisher = {American Physical Society},
  doi = {10.1103/PhysRevB.108.L140505},
  urldate = {2025-02-24},
}

@article{2025Wangl,
  title = {Theory of {{Pressure Dependence}} of {{Superconductivity}} in {{Bilayer Nickelate}} \$\textbraceleft\textbackslash mathrm\textbraceleft{{La}}\textbraceright\textbraceright\_\textbraceleft 3\textbraceright\textbraceleft\textbackslash mathrm\textbraceleft{{Ni}}\textbraceright\textbraceright\_\textbraceleft 2\textbraceright\textbraceleft\textbackslash mathrm\textbraceleft{{O}}\textbraceright\textbraceright\_\textbraceleft 7\textbraceright\$},
  author = {Jiang, Kai-Yue and Cao, Yu-Han and Yang, Qing-Geng and Lu, Hong-Yan and Wang, Qiang-Hua},
  year = 2025,
  month = feb,
  journal = {Physical Review Letters},
  volume = {134},
  number = {7},
  pages = {076001},
  publisher = {American Physical Society},
  doi = {10.1103/PhysRevLett.134.076001},
  urldate = {2026-04-01},
}

@article{2025Xianga,
  title = {Superconductivity in Nickelate and Cuprate Superconductors with Strong Bilayer Coupling},
  author = {Fan, Zhen and Zhang, Jian-Feng and Zhan, Bo and Lv, Dingshun and Jiang, Xing-Yu and Normand, Bruce and Xiang, Tao},
  year = 2024,
  month = jul,
  journal = {Physical Review B},
  volume = {110},
  number = {2},
  pages = {024514},
  publisher = {American Physical Society},
  doi = {10.1103/PhysRevB.110.024514},
  urldate = {2025-02-24},
}

\clearpage

\appendix

\section{Generalized spin and density operators}\label{app:spin_notation}

In this section, we clarify the notation used for the generalized spin and density operators in the main text and establish their connection to the widely used orbital pseudospin formalism in the literature on spin-orbital physics~\cite{1973KK}.

The local operators in $\mathcal{H}_{\rm ex}$~\eqref{eq:Hex} are constructed from the generalized density and spin operators, defined as:
\begin{align}
    n_{i,\alpha\beta} &= c_{i\alpha}^\dagger c_{i\beta} = \sum_{s} c_{i\alpha s}^\dagger c_{i\beta s} \label{eq:app_n_def} \\
    \mathbf{S}_{i,\alpha\beta} &= \frac{1}{2} c_{i\alpha}^\dagger \bm{\sigma} c_{i\beta} = \frac{1}{2} \sum_{ss'} c_{i\alpha s}^\dagger \bm{\sigma}_{ss'} c_{i\beta s'} \label{eq:app_s_def}
\end{align}
where $c_{i\alpha}^\dagger = (c^\dagger_{i\alpha\uparrow}, c^\dagger_{i\alpha\downarrow})$ is a spinor and $\bm{\sigma}$ is the vector of Pauli matrices. This notation is particularly convenient as it arises directly from the second-order perturbation expansion of the two-orbital Hubbard model, where interactions naturally couple states via the matrix elements $t_{ij}^{\alpha\beta}$.

In many theoretical treatments, it is common to introduce a set of local operators based on a pseudospin-$\frac{1}{2}$ representation for the orbital degree of freedom. Let us define the orbital pseudospin operators $\bm{\tau}$, where $\tau^a$ ($a=x,y,z$) are three Pauli matrices acting on the orbital space $\{\alpha=0, \alpha=1\}$.
Then, the 16 operators formed by the tensor product $\{\tau^a \otimes \sigma^b\}$, where $a,b \in \{0, x, y, z\}$ ($\tau^0$ and $\sigma^0$ are identity matrices), form a complete basis for all local operators acting on the spin-orbital Hilbert space of a single site.

\begin{table}[h]
\centering
\renewcommand{\arraystretch}{1.3}
\caption{Correspondence between the generalized operators ($n_{i,\alpha\beta}, \mathbf{S}_{i,\alpha\beta}$) and the standard physical operators expressed in the orbital pseudospin formalism. Here we introduce the four-component fermion vector $\Psi_i^\dagger = (c_{i0\uparrow}^\dagger, c_{i0\downarrow}^\dagger, c_{i1\uparrow}^\dagger, c_{i1\downarrow}^\dagger)$ at site $i$.}
\label{tab:operator_correspondence}
\begin{tabular}{lll}
\hline
\hline
\textbf{Pseudospin} & $\qquad\qquad$ &  \textbf{Generalized spin/density} \\
\hline
$n_i=\Psi_i^\dagger (\tau^0 \otimes \sigma^0) \Psi_i$ & & $n_{i,00} + n_{i,11}$ \\
$S^a_i=\Psi_i^\dagger(\tau^0 \otimes \frac{\sigma^a}{2}) \Psi_i$ & & $S^a_{i,00} + S^a_{i,11}$ \\
$T_i^z=\Psi_i^\dagger(\frac{\tau^z}{2} \otimes \sigma^0)\Psi_i$ & &$\displaystyle (n_{i,00} - n_{i,11})/2$ \\
 $T_i^x=\Psi_i^\dagger(\frac{\tau^x}{2} \otimes \sigma^0)\Psi_i$ & &$(n_{i,01} + n_{i,10})/2$ \\
 $T_i^y=\Psi_i^\dagger(\frac{\tau^y}{2} \otimes \sigma^0)\Psi_i$ & &$(n_{i,01} - n_{i,10})/2i$ \\
 $\Psi_i^\dagger (\tau^z \otimes \frac{\sigma^a}{2}) \Psi_i$ & & $S^a_{i,00} - S^a_{i,11}$ \\
 $\Psi_i^\dagger (\tau^x \otimes \frac{\sigma^a}{2}) \Psi_i$ & & $S^a_{i,01} + S^a_{i,10}$ \\
 $\Psi_i^\dagger (\tau^y \otimes \frac{\sigma^a}{2}) \Psi_i$ & &$i(S^a_{i,10} - S^a_{i,01})$ \\
\hline
\hline
\end{tabular}
\end{table}

The connection between our generalized operators and the standard physical observables expressed in the pseudospin formalism is listed in Table.~\ref{tab:operator_correspondence}. 
In summary, while the pseudospin formalism is elegant for describing possible SU(4) symmetries, the generalized operators $n_{i,\alpha\beta}$ and $\mathbf{S}_{i,\alpha\beta}$ used in our work provide the most direct and compact representation of the exchange Hamiltonian derived from perturbation theory.

\section{Variational Monte Carlo methods}\label{methods}

The variational ground state is constructed using the Gutzwiller-projected wave function ansatz:
\begin{equation}
    |\psi\rangle=P_G|\psi_{MF}\rangle,
\end{equation}
where $|\psi_{\text{MF}}\rangle$ is a free-fermion state, and $P_G$ is the Gutzwiller projector that strictly enforces the single-occupancy constraint at each site $i$: $$\sum_{\alpha=0,1}\sum_{\sigma=\uparrow,\downarrow} n_{i\alpha\sigma} \le 1.$$         

The free-fermion state $|\psi_{\text{MF}}\rangle$ is the ground state of a general trial mean-field Hamiltonian:
\begin{align}
    H_{MF}=\sum_{\langle ij\rangle,\sigma}\sum_{\alpha\beta} -\chi_{ij}^{\alpha\beta} (c_{i\alpha\sigma}^\dagger c_{j\beta\sigma} + \text{h.c.})-\mu\sum_{i\alpha\sigma} n_{i\alpha\sigma} \notag\\
    + \sum_{\langle ij\rangle}\sum_{\alpha\beta}\left[-\eta_{ij}^{\alpha\beta}(c_{i\alpha\uparrow}^\dagger c_{j\beta\downarrow}^\dagger - c_{i\alpha\downarrow}^\dagger c_{j\beta\uparrow}^\dagger)+\text{h.c.}\right].
    \label{eq:HMF}
\end{align}
Here, $\chi_{ij}^{\alpha\beta}$ and $\eta_{ij}^{\alpha\beta}$ are complex variational parameters representing the renormalized hopping amplitudes and superconducting pairing strengths, respectively, and $\mu$ is the variational chemical potential. We assume singlet pairing with orbital symmetry $\eta_{ij}^{\alpha\beta} = \eta_{ij}^{\beta\alpha}$.
To reduce the number of independent parameters, we impose translational invariance and the point-group symmetries of the underlying lattice. In this way, each parameter on a square lattice can take two values, depending on the bond directions, namely $\chi_x^{\alpha\beta}, \chi_y^{\alpha\beta}$ for hoppings and $\eta_x^{\alpha\beta}, \eta_y^{\alpha\beta}$ for pairings. The hopping parameters follow the sign of the hopping amplitudes in Eq.~\eqref{eq:Hkn} with
\begin{equation}
\chi_x^{\alpha\alpha}=\chi_y^{\alpha\alpha}\equiv\chi^{\alpha\alpha},\quad \chi_{x}^{01}=\chi_{x}^{10}=-\chi_{y}^{01}=-\chi_{y}^{10}\equiv\chi^{01}.
\end{equation}

For the pairing symmetry, we investigate both extended $s$-wave and $d$-wave channels. The corresponding constraints on the gap parameters are defined as:
\begin{align}
    d\text{-wave}: & \quad \eta_{x}^{\alpha\alpha} = -\eta_y^{\alpha\alpha}, \quad \eta_{x}^{01} = \eta_y^{01}, \\
    s\text{-wave}: & \quad \eta_{x}^{\alpha\alpha} = \eta_y^{\alpha\alpha}, \quad \eta_{x}^{01} = -\eta_y^{01}.
\end{align}
We note that preliminary calculations indicate the $d$-wave symmetry is energetically favorable compared to the $s$-wave and other competing ans\"atze.

The optimal ground state is determined by minimizing the total energy expectation value $E = \langle \psi | \mathcal{H} | \psi \rangle / \langle \psi | \psi \rangle$ with respect to the variational parameter set $\{\chi_{ij}^{\alpha\beta}, \eta_{ij}^{\alpha\beta}, \mu\}$. The optimization is performed on an $L \times L$ lattice with periodic boundary conditions using the standard stochastic-reconfiguration method; see details in Refs.\cite{vmc_sorella_1,vmc_sorella_2}. For a given model parameter, calculations are initialized with various configurations to avoid local minima.

\end{document}